\documentclass{PoS}

\title{The RPC-based proposal for the ATLAS forward muon trigger upgrade in view of super-LHC}

\ShortTitle{The RPC-based proposal for the ATLAS forward muon trigger upgrade in view of super-LHC}

\author{\speaker{Junjie Zhu}\\
        University of Michigan, Ann Arbor, MI, 48109\\
       On behalf of the ATLAS Muon Collaboration\\
        E-mail: \email{junjie@umich.edu}}

\abstract{The innermost station of the present ATLAS forward muon detector needs to be upgraded for the super-LHC. 
We present a  proposal to replace it with a sandwiched detector composed of several layers of small-radius Monitored Drift Tube 
chambers (sMDT) for precision tracking measurement and two stations of multi-gap Resistive Plate Chambers (mRPC) for 
triggering purpose. We describe the layout of the upgraded detector and the trigger strategy. 
Several modifications to the RPCs used in the ATLAS barrel region are needed to satisfy the super-LHC 
requirements. Various studies with the proposed mRPC timing resolution, spatial resolution and rate capability have been performed. }

\FullConference{XI workshop on Resistive Plate Chambers and Related Detectors - RCP2012,\\
		February 5-10, 2012\\
		INFN Laboratori Nazionali di Frascati Italy}

\begin{document}

\section{The ATLAS muon detector and muon trigger system}
The ATLAS detector~\cite{atlas_detector} is a general-purpose particle detector and 
is specifically designed to maximize the potential to uncover signs of new
physics at the LHC. A unique feature of the ATLAS detector is its large muon spectrometer 
which has a total coverage area of 5500 m$^2$~\cite{muon_detector}. 
Due to its unprecedented size, the spectrometer can detect
muons in the pseudorapidity range $|\eta| < 2.7$, and measure their momenta with a standalone transverse
momentum ($p_T$) resolution of approximately 10\% for 1 TeV muons. 
With its large acceptance and good momentum resolution, the ATLAS muon spectrometer 
has a great potential for discovery of new physics. 

\begin{figure}[!htb]
  \centering
  \includegraphics[width=0.43\textwidth,height=0.28\textwidth]{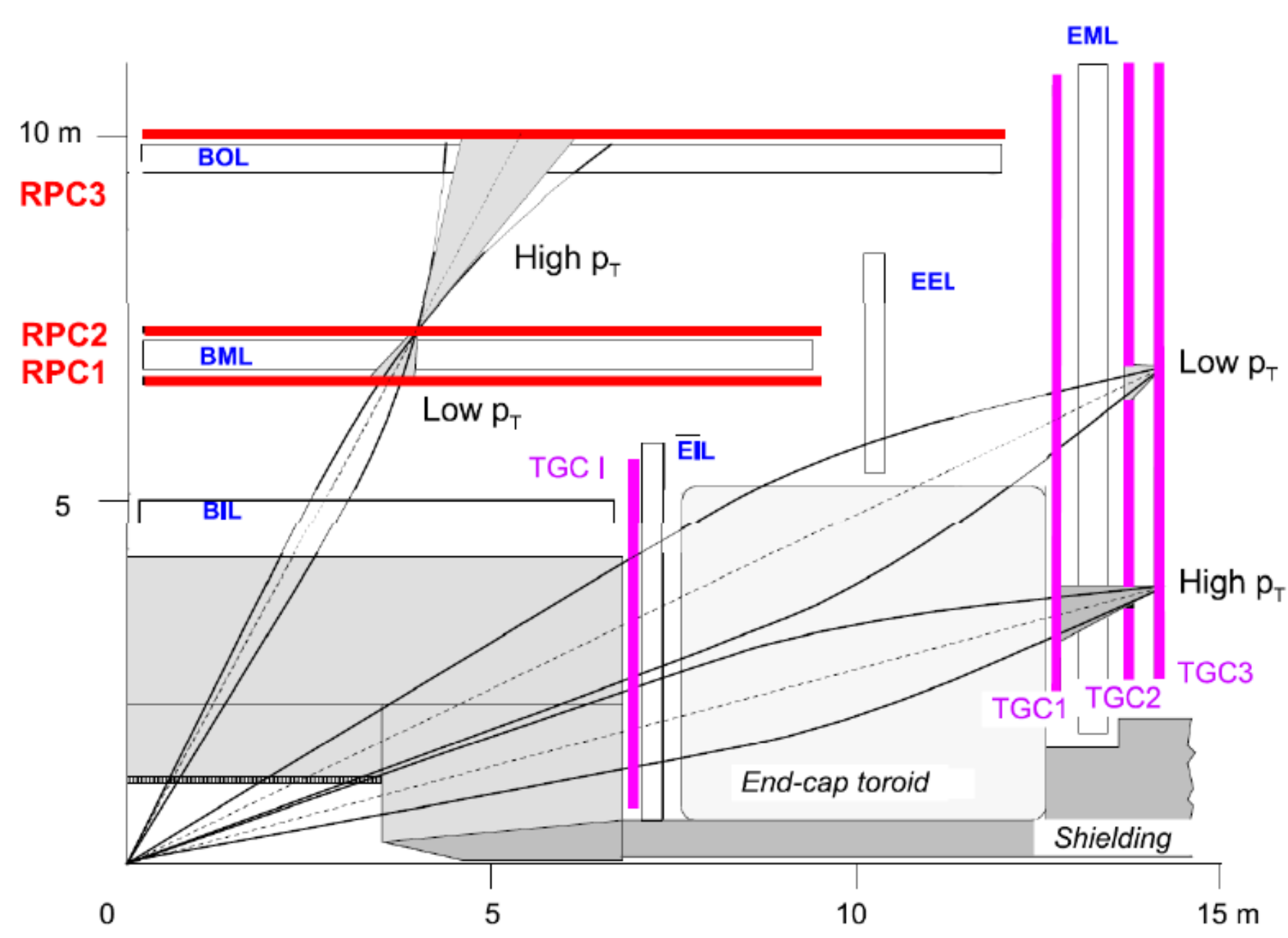}
  \includegraphics[width=0.43\textwidth,height=0.28\textwidth]{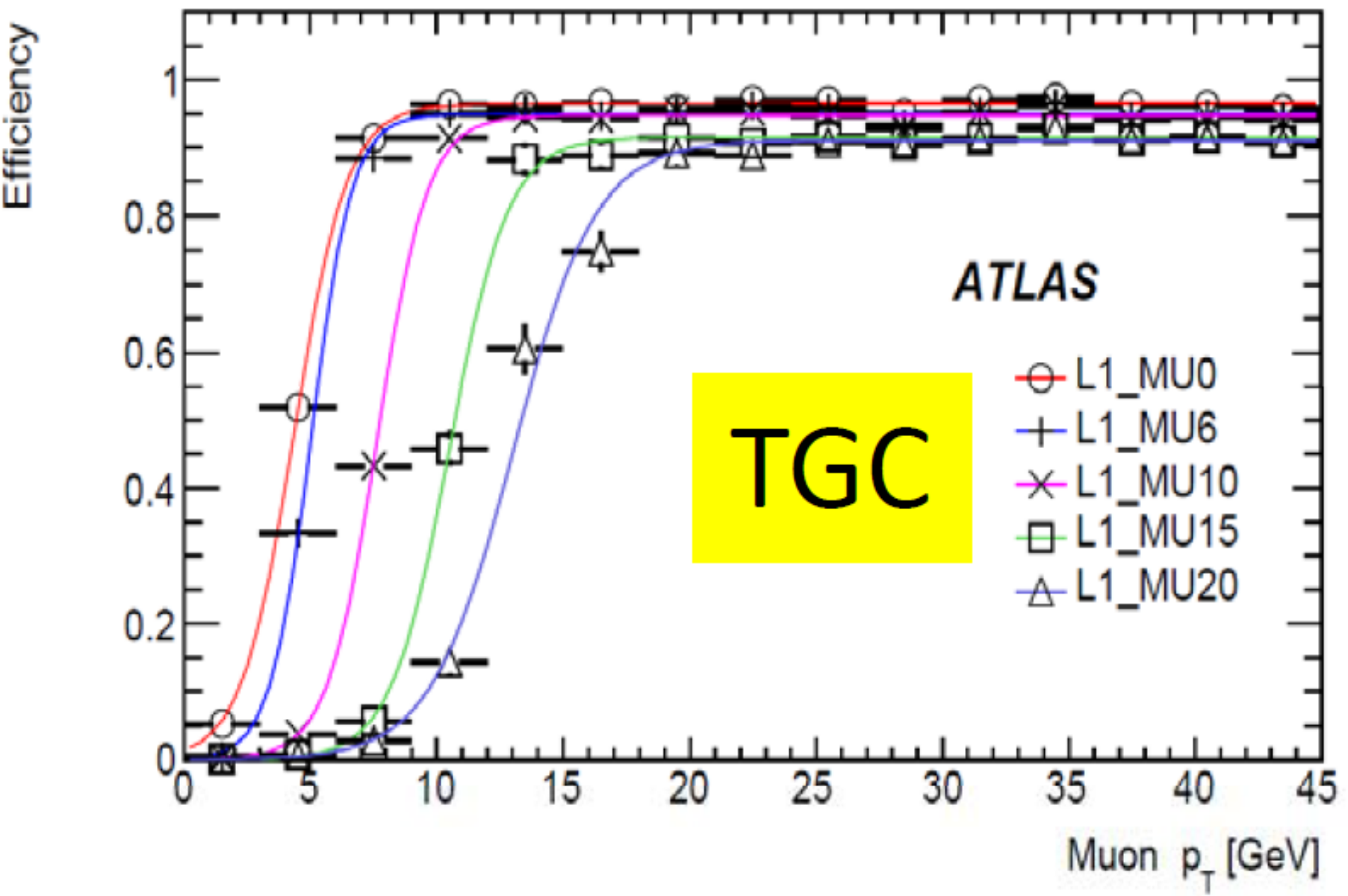}
  \caption{Left: layout of the ATLAS detector and the schematics of the muon trigger system~\cite{atlas_detector}. RPC2 and TGC3 are the reference planes for barrel 
and endcap, respectively.  Right: muon trigger efficency as a function of muon $p_T$ for five different trigger thresholds in the endcap region.} 
\label{fig:atlas_detector}
\end{figure}

A quarter view of the ATLAS muon detector is shown in the left plot of Fig.~\ref{fig:atlas_detector}. 
The precise momentum measurement is performed by 
the Monitored Drift Tube Chambers (MDT). They cover the pseudorapidity range $|\eta|<2.7$ except the forward region 
$2<|\eta|<2.7$ of the  innermost endcap layer covered by the Cathode-Strip Chambers (CSC). 
Two different technologies have been used for the trigger chambers:  the Resistive Plate Chambers (RPC) in the barrel 
region $|\eta|<1.1$ and the Thin Gap Chambers (TGC) in the endcap region $1.1<|\eta|<2.4$. A system of three 
large air-core toroids (each consists of eight coils) generates the magnetic field for the  spectrometer.

The schematic layout of the ATLAS muon trigger detector is shown also on the left plot of Fig.~\ref{fig:atlas_detector}. In the barrel, two layers (RPC1 and RPC2) sandwich the MDTs of the middle layer, while the third one (RPC3) is located close to the outer MDT layer. 
In the endcap, the three layers are in front (TGC1) and behind (TGC2 and TGC3) the second MDT wheel (called "Big Wheel (BW)"), 
while the fourth layer (TGC I) is located in front of the innermost tracking layer (called "Small Wheel (SW)"). The trigger information 
is generated by a system of fast coincidences between the three last layers along the trajectory of the muon. The deviation 
from straightness is the deviation of the slope of the track segment between two trigger chambers from the slope of a 
straight line between the interaction point and the hit in a reference layer. 

The right plot of Fig.~\ref{fig:atlas_detector} shows the L1 trigger efficiency as a function of muon $p_T$ in the endcap region 
for five different trigger thresholds (0 GeV, 6 GeV, 10 GeV, 15 GeV and 20 GeV). The plateau efficiencies for three low $p_T$ 
triggers are close to 95\% due to the coverage of the TGC detector; the plateau efficiencies for two high $p_T$ triggers 
are close to 90\% due to the requirement of  hits also in TGC1. 

\section{Problems with the current ATLAS muon triggers and future upgrade strategy}
\begin{figure}[!htb]
  \centering
  \includegraphics[width=0.43\textwidth,height=0.28\textwidth]{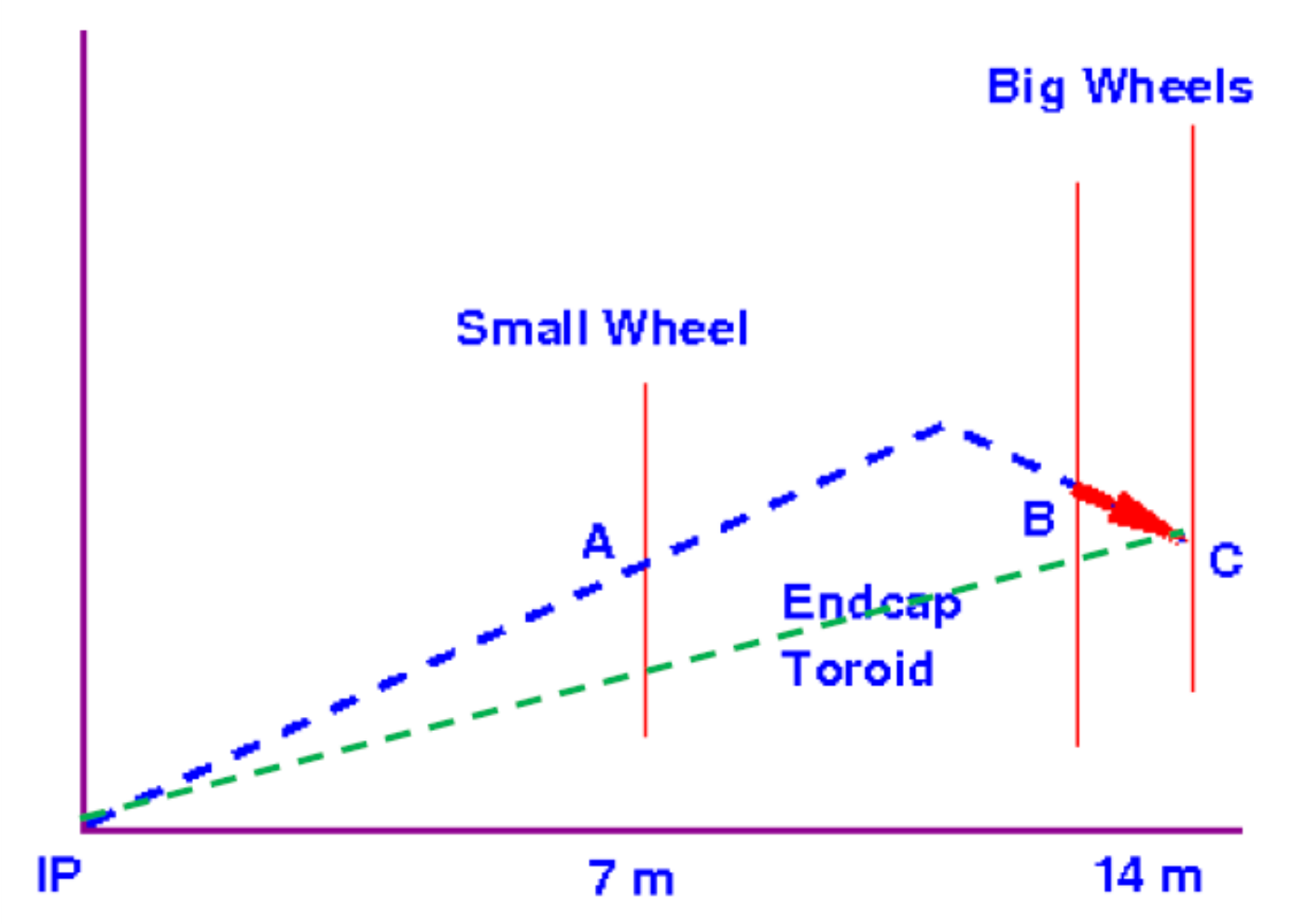}
  \includegraphics[width=0.43\textwidth,height=0.28\textwidth]{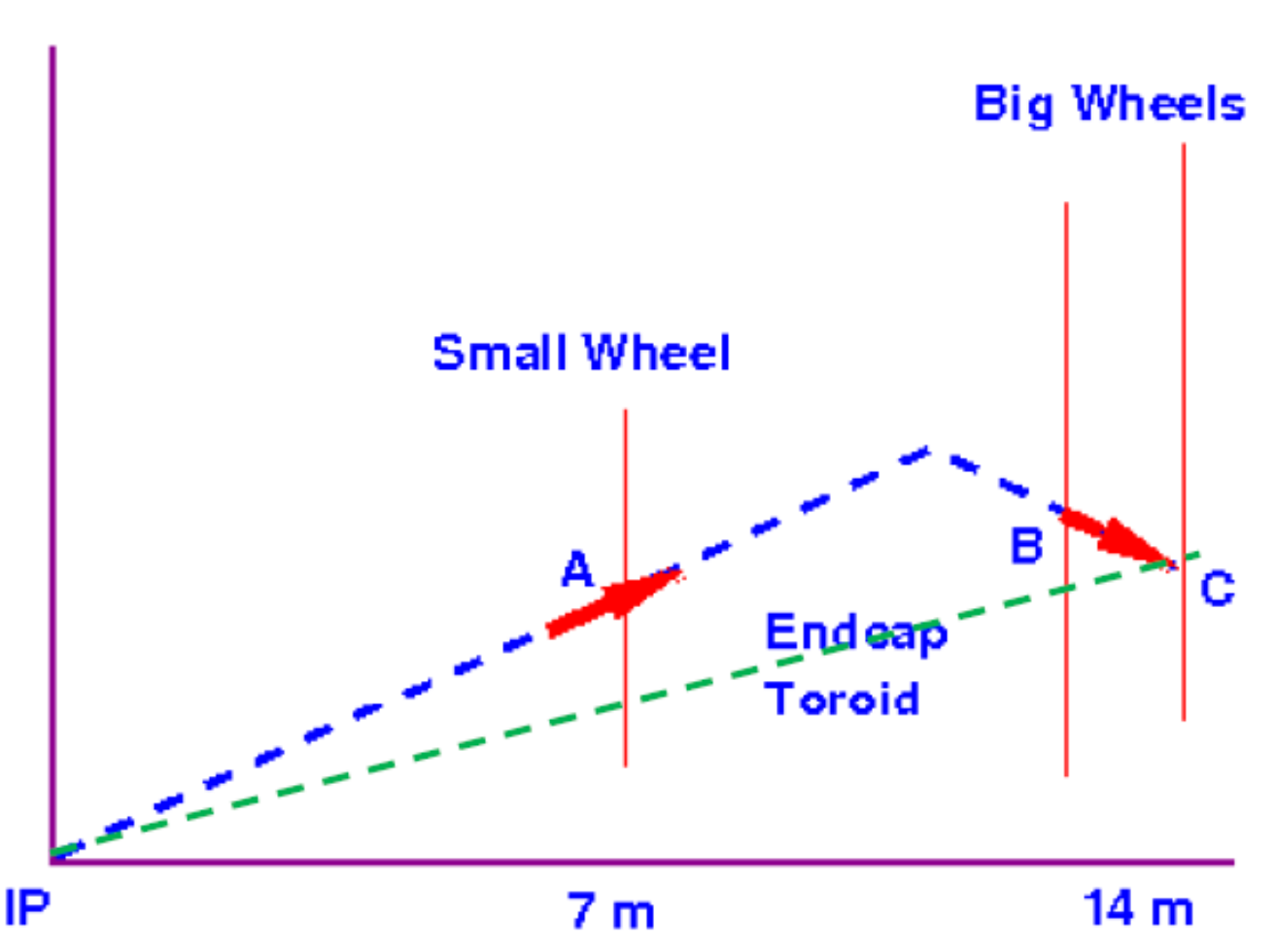}
  \caption{Left: muon trigger strategy for the present ATLAS muon detector in the endcap region; Right: muon trigger strategy with the upgraded SW detector.} 
\label{fig:trigger_strategy}
\end{figure}

The current muon trigger strategy in the endcap region  is based on the muon hits from two (or three) TGC layers at the BW. 
It is also assumed that all muons originate from the center of the detector (a LHC beam width of 5 cm along the $z$ axis is ignored). 
The basic trigger principle is shown in the left plot of Fig.~\ref{fig:trigger_strategy}, 
only a vector $BC$ at the BW is measured. Random background tracks due to other charged particles 
bent by the toroids and other neutral particles such as photons and slow neutrons 
can easily fake this condition. Studies based on the data taken in 2010 and 2011 runs indicated the following problems: 
(1) for events triggered by L1\_MU20 triggers, about 90\% of them are triggered by TGC triggers and only 10\% are triggered by RPC triggers;
(2) more than 95\% of muons triggered by the L1\_MU20 endcap trigger have no matching MDT segments in the SW; 
(3) the momentum resolution for the TGC triggers is close to 30\% for 20 GeV muons at L1 due to the worse spatial resolution of 
the TGC chambers (about a few centimeters). 
Due to these problems the trigger bandwidth is limited and the high $p_T$ muon trigger is less efficient at high luminosity. 
The problems with large fraction of fake muons and worse momentum resolution limit the trigger bandwidth and make it less 
efficient to trigger on high $p_T$ muons at high luminosity runs.  The L1 trigger rate for an instantaneous luminosity 
of $3 \times 10^{34}$ cm$^{-2}$s$^{-1}$ at the super-LHC condition is estimated to be about 60 kHz for L1\_MU20 endcap triggers. 

An upgrade project to replace the current SW tracking and trigger chambers with a new SW detector for the Phase-I 
upgrade has been agreed within the ATLAS collaboration. 
One major purpose of this upgrade is to improve the L1 muon $p_T$ resolution and remove fake muon tracks seen
by the BW TGCs at high instantaneous luminosity. 
This upgraded detector will provide a new vector $A$ (as shown in the right plot of Fig.~\ref{fig:trigger_strategy}), the difference
between the two angles measured by the new SW detector (vector $A$) and the current BW TGCs (vector $BC$) is then used to
determine the muon $p_T$ at L1. To achieve a $p_T$ resolution of 15\% for 20 GeV muons, the angular resolution provided
by the new SW detector needs to be 1 mrad. There are currently three proposals within the ATLAS muon community to replace the present
SW detector with either a Micromegas detector, a sandwich detector with small radius MDTs
(sMDT) plus either multi-gap RPCs or TGCs with finer strips. The technology to use for the upgrade
will be decided later. With an upgraded SW detector satisfying the above requirements, the L1 muon trigger rate is 
expected to be about 20 kHz for L1\_MU20 endcap triggers at an instantaneous luminosity of $3 \times 10^{34}$ cm$^{-2}$s$^{-1}$. 

\section{RPC-based proposal for the ATLAS SW upgrade}
\begin{figure}[!htb]
  \centering
  \includegraphics[width=0.43\textwidth,height=0.26\textwidth]{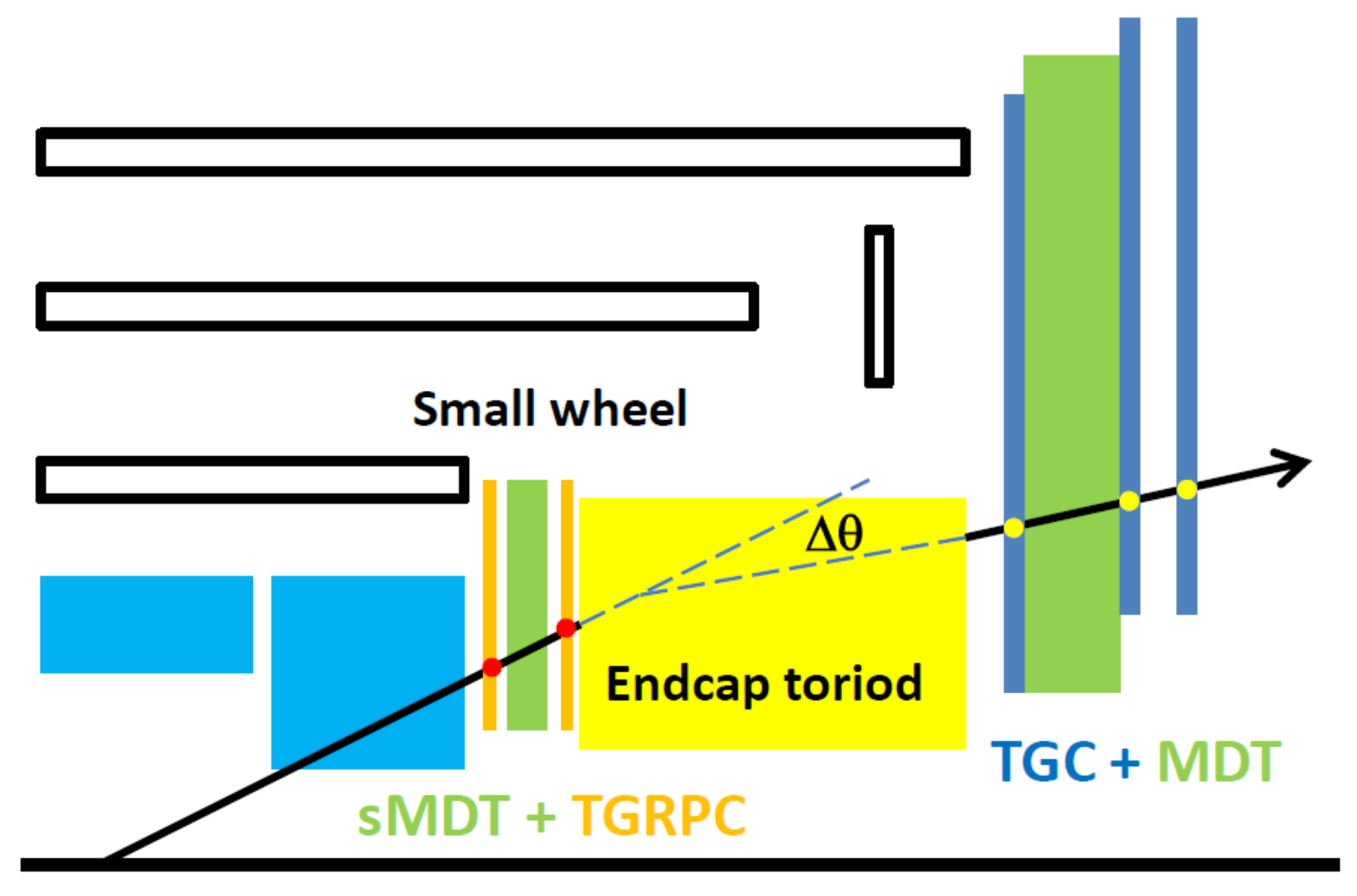}
  \includegraphics[width=0.55\textwidth,height=0.26\textwidth]{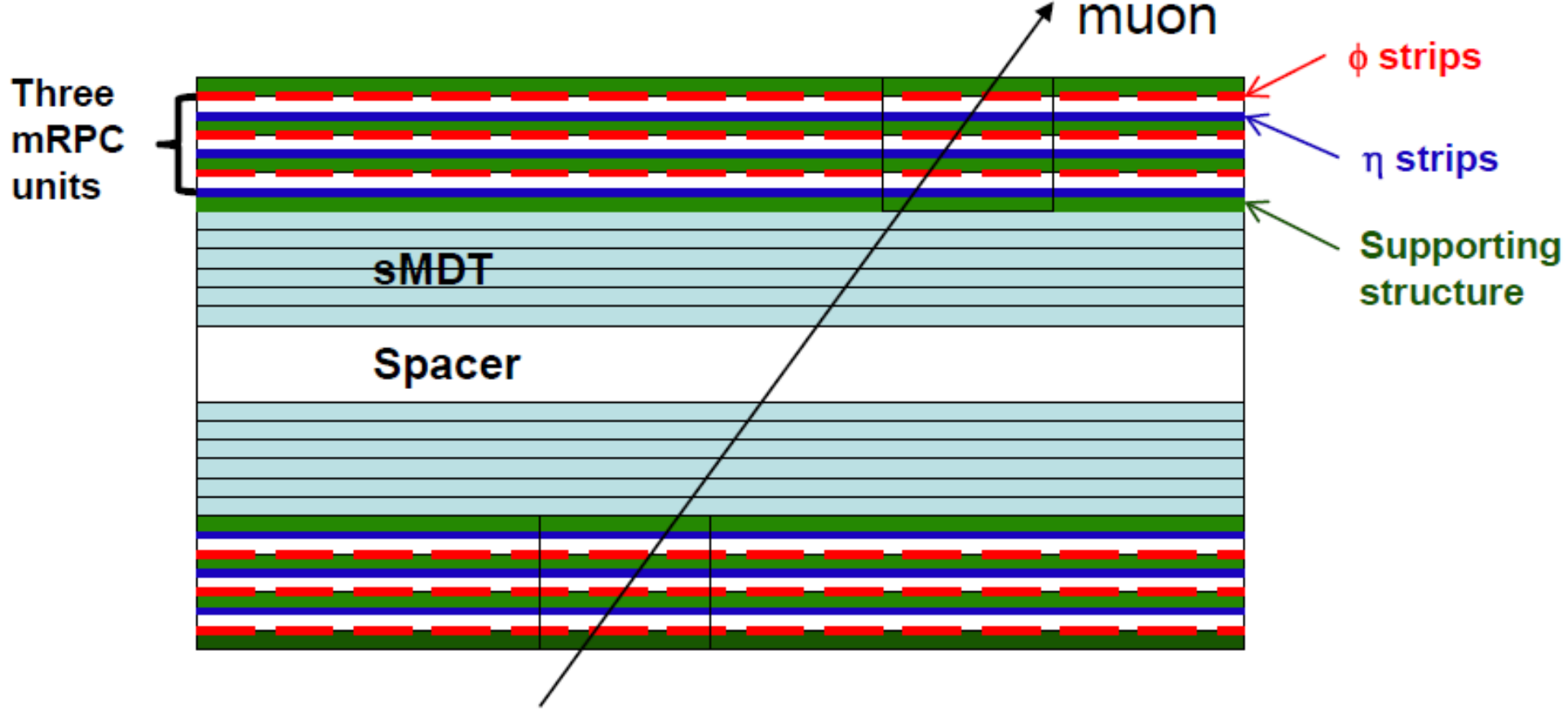}
  \caption{Left: layout of the proposed sandwiched sMDT and mRPC detector; Right: layout of sMDT $+$ six layers of mRPC detector.} 
\label{fig:layout}
\end{figure}

One proposal to upgrade the SW detector is to replace it with a sandwich of sMDT for precision tracking 
and two stations of multi-gap RPCs (mRPC) for triggering~\cite{upgrade}. The layout of the new SW detector is shown in the 
left plot of Fig.~\ref{fig:layout}.  The mRPC will be assembled together with a sMDT of equal dimensions in 
a common mechanical support structure which guarantees the relative alignment of the RPC to the rest 
muon sub-detectors in the endcap region. 

One of the major reasons to use the mRPCs for the new SW detector is due to its excellent timing capability. Fast 
detectors are crucial to reject low energy uncorrelated background. 
{\sc geant}-based Monte Carlo simulation studies indicate that after the subtraction of time-of-flight, more than 90\% of the muons 
arrive within the first 2 ns, while other charged particles and neutral particles such as photons and neutrons have flatter 
arrival time distributions. For example, only about 7$-$10\% of other charged particles,  photons or neutrons arrive in the first 2 ns timing window.  
Both muons and other charged particles create correlated hits in the SW and BW detectors, while photons and neutrons create uncorrelated 
hits in the SW and BW detectors. At an uncorrelated hit rate as high as 14 kHz/cm$^2$, a very short time 
(about 2 ns) coincidence between contiguous detectors is an extremely efficient method to eliminate uncorrelated hits. 
The basic principle of the trigger scheme we propose is to remove fake muon tracks as soon as possible and as much as possible, and then use optical fibers to 
send the track hits from the SW mRPC to the trigger logic at USA15 to combine with hits provided by the BW TGCs. 

The present ATLAS  RPCs in the barrel region are based on a 2 mm gas gap between two resistive electrodes 2 mm thick, 
made of a melamine coated phenolic laminate. The gas gap is sandwiched between 
two read out panels with mutually orthogonal strips giving a point in the space for each avalanche generated inside the gas gap. 
In order to fulfill the requirements defined for the SW upgrade, a substantial improvement with respect to 
the RPCs presently operating in ATLAS is needed concerning the rate capability, time resolution, position resolution and detector ageing.
The new detector should be able to stand a rate of 14 kHz/cm$^2$ for the region closest to the beam pipe. 
The trigger decision provided by the new SW detector and BW TGCs should be available within a 25 ns LHC bunch crossing time.
In order to reach the designed angular resolution of 1 mrad, the spatial position resolution of the new detector 
should be about 300 $\mu m$ at L1. 
The total integrated charge should be around 3 C/cm$^2$ for a total integrated luminosity of 3000 fb$^{-1}$.

To increase the rate capability and reduce the ageing effect, 
two parameters have to be optimized: the gas gap width which determines the amount of delivered charge per avalanche 
and the sensitivity of front end electronics which determines the minimum charge that can be discriminated from the noise.
To obtain the required spatial resolution, we propose to use narrow readout strips, with a typical pitch of about 2 mm, 
coupled to an electronics circuit that can select the centroid position within a resolution of 300 $\mu m$ at L1.
The time required for the offline charge centroid method (charge collection and centroid calculation) would be too long at L1.  

We propose to split a single 2 mm gas gap into two 1 mm gaps giving the advantages of a better time resolution, a lower 
delivered charge which increases the rate capability. We will refer to this type of RPC as mRPC.
The electrode thickness is reduced to 1 mm to avoid large voltage drop on the electrode plates due to large current expected 
at high instantaneous luminosity. 
A typical 2 mm strip pitch is expected for the readout strips on the bending plane. 
The strips will be equipped with both ends connected to a mean-timer electronics circuit. 
The hit position determined using time difference for the signals from the two ends will be 
localized along the strip with a typical uncertainty of a few cm 
giving a further reduction of uncorrelated background in the coincidence of two contiguous chambers. The probability for 
two or more muons crossing the same readout strip is estimated to be below 0.01\%. 
To obtain the centroid of the hits, we can either explore the fact that strips with larger charge deposition will cross 
the thresholds earlier using the mean-timer circuit or group 8 $-$ 10 readout strips in super-strips connected 
to a maximum selector which identifies, in a typical time of 10 ns, the strip with the maximum charge deposition.

The layout of a proposed sMDT plus mRPC unit is shown in Fig.~\ref{fig:layout}. Three layers of mRPC detectors are arranged 
on both sides of an sMDT chamber. Short time coincidences among detector layers of the same triplet can strongly 
reduce the uncorrelated background. The trigger logic requires the 2 out of 3 coincidence inside the 
same triplet followed by the twofold coincidence of the two triplet signals to remove almost all uncorrelated hits and still retain high efficiency. 
To reduce correlated hits produced by other charged particles, the angle provided by the two mRPC stations is required to be consistent with 
the angle for muons originating from the center of the detector.
 
The proposed trigger strategy has the following advantages: (1) obtains unambiguous identification of bunch  crossing ID; 
(2) provides simple on-chamber pattern recognition; (3) sends small amount of information to combine with track segments 
from the BW TGCs; (4) performs the coincidence between the SW and BW in the counting room; and (5) provides 
significant safety margin. 

\section{Studies with the RPC}
To examine the mean-timer technique we have employed the MDT electronics for readout of
RPC signals from both ends of the signal lines. Although these electronics are not adequately fast for a
final system evaluation, it is fast enough to assess off-line the algorithms proposed. 
Studies indicate that a timing resolution of 0.6 ns is obtained for a typical RPC strip. 
The features of the MDT front-end that limit its timing precision are the slow peaking times (15 ns), 
the channel to channel threshold variations, and the 0.78 ns measurement bin. We are performing further tests with 
faster electronics developed by the ALICE experiment and we expect a timing resolution of 15 $-$ 100 ps that can give a position 
resolution of 1 $-$ 2 cm along the strip. We are also investigating sensitive front end readout electronics that can select smaller signals due to smaller gaps~\cite{roberto_talk}.

To study the RPC spatial resolution, we used a single-gap glass RPC chamber with a thickness of 0.6 $-$ 0.8 mm for the glass plates 
and a thickness of 1.2 mm for the gas gap. The readout strips have a pitch size of 1.27 mm and were readout at both ends by the 
present ATLAS MDT front-end readout electronics. Studies were done with 180 GeV muon beam at the CERN H8. The RPC chamber was placed in 
front of twelve layers of sMDTs (with a radius of 1.5 cm). The difference between the hit position determined by the RPC chamber compared 
to the extrapolated position from the sMDT chamber indicates the spatial resolution of the RPC chamber. 
A spatial resolution of $288 \pm 7$ $\mu m$ was obtained using the offline charge centroid method and $374 \pm 11$ $\mu m$ using 
the timing information from the  mean-timer circuit. More than 100 $\mu m$ systematic 
uncertainty was expected from the sMDT spatial resolution and relative alignment between these two detectors. 

\begin{figure}[!htb]
  \centering
  \includegraphics[width=0.43\textwidth,height=0.26\textwidth]{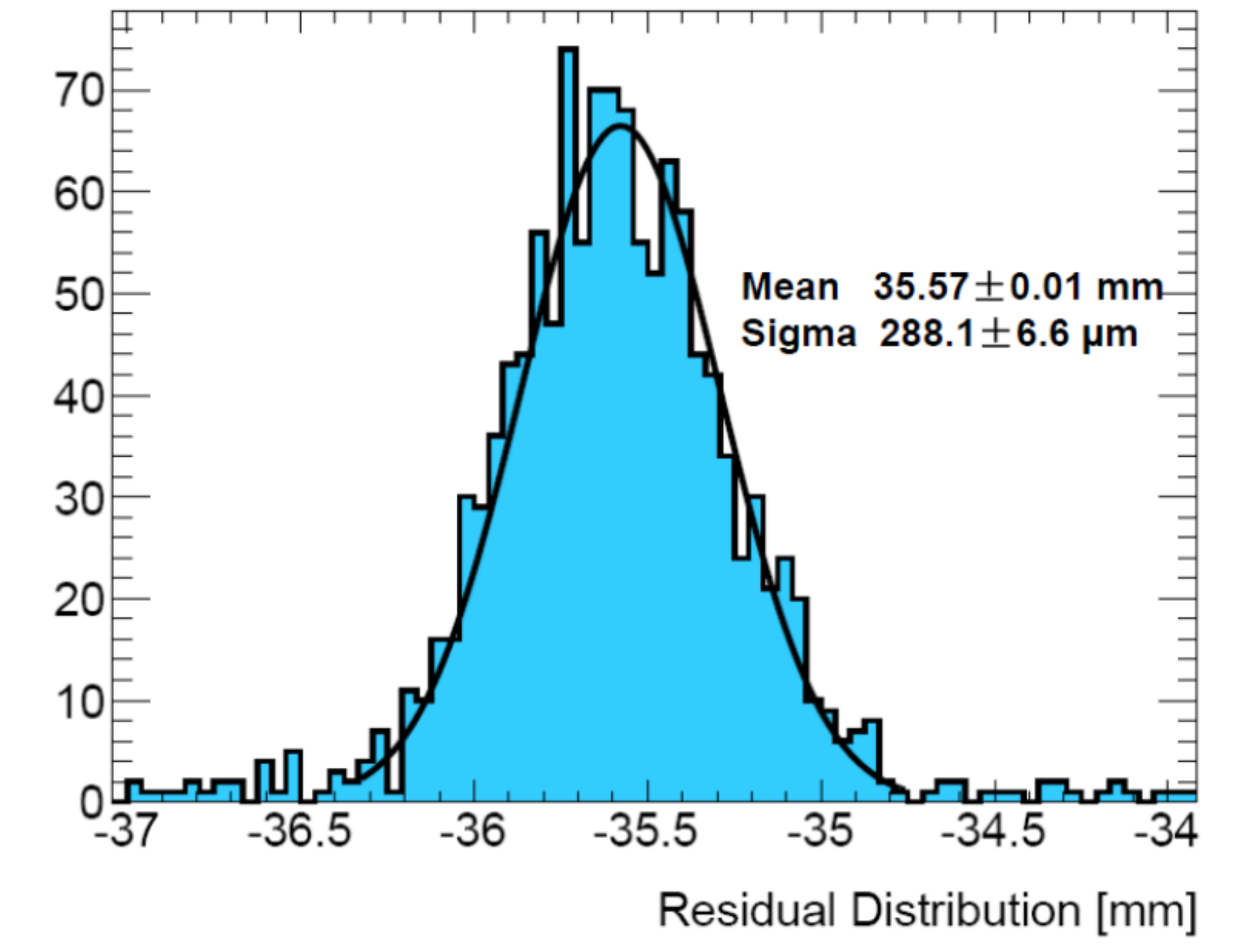}
  \includegraphics[width=0.55\textwidth,height=0.26\textwidth]{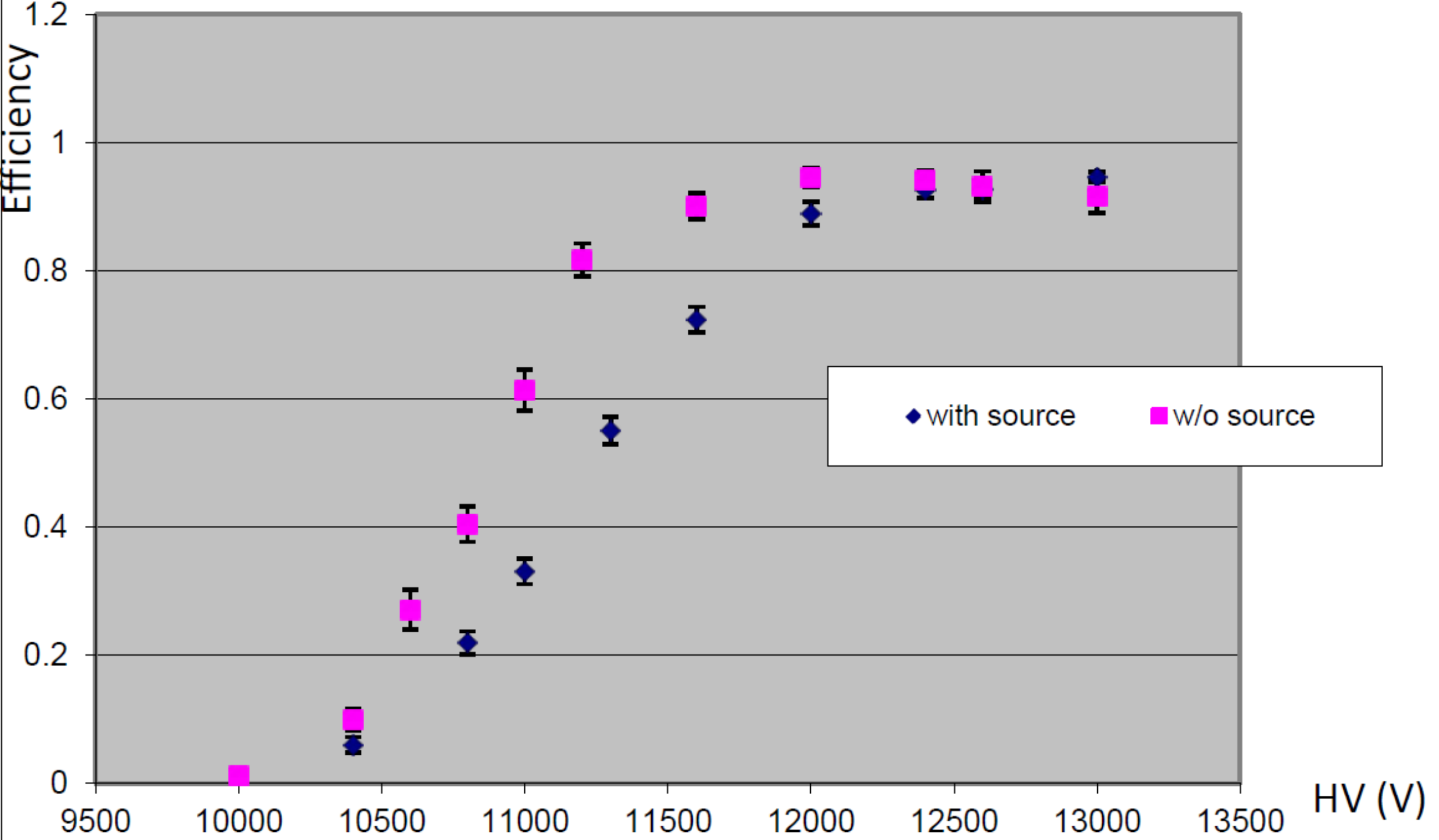}
  \caption{Left: spatial resolution using the charge-centroild method; Right: mRPC detection efficiency as a function of the applied high voltage with and without the radiative source.} 
\label{fig:test}
\end{figure}

The rate capability studies were performed using a mRPC with two gas gaps (2 mm bakelite plates and 1 mm gas gap) that is placed close to 
a $Cs^{137}$ source at the CERN GIF. The cosmic muon detection efficiency as a function of the high voltage applied is shown in Fig.~\ref{fig:test}
with and without the radiative source nearby. Similar detection efficiency was achieved at 12.5 kV even though there was a shift of a few hundred voltage for 
these two situations. The rate capability test was performed at a rate of 7 kHz/cm$^2$ and is limited by the available source flux at GIF. 
 
\section{Conclusion}
We propose to upgrade the ATLAS SW muon detector with a sandwiched detector composed of sMDT for precision tracking 
and two stations of mRPC for triggering. We made several modifications to the present ATLAS RPCs used 
in the barrel region and have perforned studies with the timing resolution, position resolution and rate capability.

\end{document}